\documentstyle[12pt]{article}

\begin{document}
\setcounter{page}{1} \pagestyle{plain} \vspace{1cm}
\begin{center}
\Large{\bf Energy conditions in $F(T,\Theta)$ gravity and compatibility with a stable de Sitter solution}\\
\small \vspace{1cm}
{\bf {\bf Faeze Kiani\footnote{fkiani@umz.ac.ir}\quad\ and \quad Kourosh Nozari\footnote{knozari@umz.ac.ir}}}\\
\vspace{0.5cm} {\it Department of Physics,
Faculty of Basic Sciences,\\
University of Mazandaran,\\
P. O. Box 47416-95447, Babolsar, IRAN}\\
\end{center}
\vspace{1.5cm}
\begin{abstract}
We study a new type of the modified teleparallel gravity of the form
$F(T,\Theta)$ in which $T$, the torsion scalar, is coupled with
$\Theta$, the trace of the stress-energy tensor. In a perturbational
approach, we study the stability of the solutions and as a special
case we find a condition for stability of the de Sitter phase. Then
we adopt a suitable form for $F(T,\Theta)$ that realizes a stable de
Sitter solution so that the stability condition creates a specific
constraint on the parametric space of the model. Finally, the energy
conditions in the framework of $F(T,\Theta)$ gravity is
investigated. 
\end{abstract}
\vspace{1.5cm}
\newpage

\section{Introduction}
Einstein's general relativity is a completely geometrical theory so
that gravitation is described not as a force, but as a geometric
deformation of flat Minkowski space-time. In this point of view, the
gravitational field creates a curvature in space time that its
action on the particles is determined by allowing them to follow the
geodesics of the space time. In this approach, trajectories is
described by the geodesic equation not the force equation [1]. On
the other hand, in 1928, Einstein in an attempt to build a unified
gauge theory of gravitation and electrodynamics presented the other
theory of gravity, the so-called teleparallel gravity [2]. In this
theory torsion, the antisymmetric part of connection, is non-zero
and torsion instead of curvature describes the gravitational
interaction. In teleparallel gravity, tetrad (or vierbein) fields
form the (pseudo) orthogonal bases for the tangent space at each
point of flat space time. Similar to the metric tensor in general
relativity, here tetrad play the role of the dynamical variables.
Teleparallel gravity also uses the curvature-free Weitzenb\"{o}ck
connection instead of Levi-Civita connection of general relativity
to define covariant derivatives [3]. In spite of the such
fundamental conceptual differences between teleparallel theory and
general relativity, it has been shown that teleparallel Lagrangian
density only differs with Ricci scalar by a total divergence
[4,5].\\

In general relativity, the dark energy puzzle can be addressed by
introducing additional geometrical degree of freedom into the
theory, that is called $F(R)$ modified gravity. In $F(R)$ gravity
the late time acceleration of the universe is catched by dark
geometry instead of dark energy [6]. The modification of gravity in
teleparallel gravity is accomplished by supplementing an additional
torsion term into Einstein-Hilbert Lagrangian [7]. The $F(T)$
gravity has interesting properties that the field equations are of
second order, unlike $F(R)$ gravity which is of fourth order in the
metric approach. In this context, $F(T)$ models have been
extensively used in cosmology to explain the late time cosmic
speed-up expansion without the need of dark energy [7,8].\\

In this paper we construct a generalization of $F(T)$ modified
gravity by considering coupling between torsion scalar $T$ and trace
of the stress-energy tensor $\Theta$ via a general function as
$F(T,\,\Theta)$. Then we investigate stability of the de Sitter
solution (when subjected to homogeneous perturbations) in this
framework.  In this sense, we obtain a stability condition for the
de Sitter phase in the general $F(T,\Theta)$ theories. Then we
propose a specific $F(T,\Theta)$ model and show that the stability
condition can be expressed as a constraint equation between the
parameters of the model. We also consider the constraints imposed by
the energy conditions and investigate whether the parameters ranges
of the proposed model are consistent with the stability conditions.
We note that since homogeneous and isotropic perturbations can be
considered as the route to determine the stability of different
modified gravity theories (see for instance [9]), the full
anisotropic analysis of the cosmological perturbations is not
considered here.

The paper is organized as follows: in Section II the general
features of the $F(T,\Theta)$ theories is explored by writing the
corresponding modified Einstein equations. In Section III the
evolution equations of corresponding perturbations in FRW background
is introduced. In section IV we devote to the study of stability
around the de Sitter solution. In section V we present the energy
conditions in $F(T,\Theta)$ gravity and compare the results with the
obtained constraints from the stability conditions. We close the
paper by giving our conclusions in Section VI.

\section{$F(T,\Theta)$ Gravity}
In this section, firstly a general $F(T,\Theta)$ function is
considered for Lagrangian density of the action as follows
\begin{equation}
S=\int{e\,\Big(\frac{F(T,\Theta)}{2\kappa^{2}}+ L_{m}\Big)\, d^{4}x}
\end{equation}
where $\kappa^{2}=8\pi G$. $e=\sqrt{-g}$ is determinant of the
vierbein $e^{i}_{\mu}$ and $T$ is the torsion scalar.  $g$ is the
determinant of the metric tensor and the metric of the space-time
$g_{\mu\nu}$ is related to the vierbein by
$g_{\mu\nu}=\eta_{ij}\,e^{i}_{\mu}\,e^{j}_{\nu}$\,. Here we use the
Greek alphabet ($\mu$, $\nu$, $\rho$, ... = 0, 1, 2, 3) to denote
indices related to spacetime, and the Latin alphabet ($i$, $j$, $k$,
... = 0, 1, 2, 3) to denote indices related to the tangent space.
The ordinary matter part of the action is shown by $L_{m}$ and the
corresponding stress-energy tensor is
\begin{equation}
\Theta_{i}^{\mu}=-2\frac{\partial L_{m}}{\partial
e^{i}_{\mu}}-2e^{\mu}_{i}\,L_{m}
\end{equation}
The connection that is used in general relativity, is the
Levi-Civita connection
\begin{equation}
\hat{\Gamma}^{\rho}_{\,\,\mu\nu}=\frac{1}{2}g^{\rho\sigma}(\partial_{\nu}g_{\sigma\mu}+\partial_{\mu}g_{\sigma\nu}-\partial_{\sigma}g_{\mu\nu})
\end{equation}
This connection leads us to nonzero spacetime curvature but zero
torsion [10]. In teleparallel gravity, tetrad fields give rise to a
connection namely the Weitzenb\"{o}ck connection, instead of the
Levi-Civita connection, which is defined by
\begin{equation}
\tilde{\Gamma}^{\lambda}_{\,\,\mu\nu}=e^{\lambda}_{i}\partial_{\nu}e^{i}_{\mu}=-e^{i}_{\mu}\partial_{\nu}e^{\lambda}_{i}
\end{equation}
One of the consequences of this definition is that the covariant
derivative, $D_{\mu}$, of the tetrad fields vanishes identically:
\begin{equation}
D_{\mu}e_{\nu}^{i}\equiv\partial_{\mu}e_{\nu}^{i}-\tilde{\Gamma}^{\lambda}_{\,\,\nu\mu}e_{\lambda}^{i}=0
\end{equation}
This equation leads us to a zero curvature but nonzero torsion [10].
We define the torsion and contortion by
\begin{equation}
T_{\,\,\mu\nu}^{\rho}=\tilde{\Gamma}_{\,\,\nu\mu}^{\rho}-\tilde{\Gamma}_{\,\,\mu\nu}^{\rho}
\end{equation}
\begin{equation}
K_{\,\,\mu\nu}^{\rho}=\tilde{\Gamma}_{\,\,\mu\nu}^{\rho}-\hat{\Gamma}_{\,\,\mu\nu}^{\rho}=\frac{1}{2}(T_{\mu\,\,\,\,\,\nu}^{\,\,\,\rho}+
T_{\nu\,\,\,\,\,\mu}^{\,\,\,\rho}-T_{\,\,\,\,\,\mu\nu}^{\rho})
\end{equation}
respectively, that the contortion is expressed as the interrelation
between Weitzenb\"{o}ck and Levi-Civita connections [4]. Now, one
can define super-potential as follows
\begin{equation}
S_{\sigma}^{\,\,\mu\nu}\equiv \frac{1}{2}\Big(
K_{\quad\sigma}^{\mu\nu}+\delta^{\mu}_{\sigma}T_{\quad\xi}^{\xi\nu}-\delta^{\nu}_{\sigma}T_{\quad\xi}^{\xi\mu}\Big)
\end{equation}
to obtain the torsion scalar
\begin{equation}
T\equiv
S_{\sigma}^{\,\,\,\mu\nu}T^{\sigma}_{\,\,\,\mu\nu}=\frac{1}{4}T^{\xi\mu\nu}T_{\xi\mu\nu}+\frac{1}{2}T^{\xi\mu\nu}T_{\nu\mu\xi}
-T_{\xi\mu}^{\quad\xi}\,T^{\nu\mu}_{\quad\nu}
\end{equation}
which is used as the Lagrangian density in formulation of the
teleparallel theories.\\
The generalized field equations are
extracted by varying of the expression (1) with respect to the
vierbein field $e_{\nu}^{i}$ as follows
$$e^{-1}\,\partial_{\rho}(e\,S_{i}^{\,\,\mu\rho})\,F_{T}+e^{-1}\,\partial_{\rho}(e\,\Xi_{i}^{\,\,\mu\rho})F_{\Theta}+
S_{i}^{\,\,\mu\rho}\,\partial_{\rho}T\,F_{TT}+\Xi_{i}^{\,\,\mu\rho}\,\partial_{\rho}\Theta\,F_{\Theta\Theta}\,+$$
\begin{equation}
\frac{1}{4}e_{i}^{\mu}F- T^{\sigma}_{\,\,\nu
i}\,S_{\sigma}^{\,\,\mu\nu}F_{T} -
\Upsilon_{i}^{\,\mu}F_{\Theta}=\,4\pi G
e_{i}^{\sigma}\,\Theta_{\sigma}^{\,\mu}
\end{equation}
where
$\Xi_{i}^{\,\mu\rho}=\partial\Theta/\partial\,e_{\mu,\rho}^{i}$ and
$\Upsilon_{i}^{\,\mu}=\frac{1}{4}\,\partial\Theta/\partial
e_{\,\mu}^{i}$. Note that $F_{T}$ and $F_{\Theta}$ ($F_{TT}$ and
$F_{\Theta\Theta}$) are the first (second) derivatives of the
$F(T,\Theta)$ with respect to $T$ and $\Theta$, respectively.\\
Here, it is assumed that the Lagrangian of matter is only in terms
of $e_{\,\mu}^{i}$, and so $\Xi_{i}^{\,\,\mu\rho}$ is zero. Since
$\Theta=e_{\alpha}^{j}\,\Theta_{j}^{\alpha}$, one can say
$\Upsilon_{i}^{\,\mu}$ is written as
$\Upsilon_{i}^{\,\mu}=\frac{1}{4}\,\Theta_{i}^{\,\mu}+\Omega_{i}^{\mu}$
that
$\Omega_{i}^{\mu}=\frac{1}{4}\,e_{\alpha}^{j}\,[\partial\Theta_{j}^{\alpha}/\partial
e_{\,\mu}^{i}]$. So the field equations (10) can be rewritten as
follows
\begin{equation}
e^{-1}\,\partial_{\rho}(e\,S_{i}^{\,\,\mu\rho})\,F_{T}+
S_{i}^{\,\,\mu\rho}\,(\partial_{\rho}T)\,F_{TT}
+\frac{1}{4}e_{i}^{\mu}F-e_{i}^{\gamma}\,T^{\sigma}_{\,\,\nu
\gamma}\,S_{\sigma}^{\,\,\mu\nu}F_{T}-\Omega_{i}^{\,\mu}\,F_{\Theta}=\,4\pi
G\,\Theta_{i}^{\,\mu}+\frac{1}{4}\,F_{\Theta}\,\Theta_{i}^{\,\mu}
\end{equation}
On the other hand, $\Omega^{\mu}_{i}$ by definition (2) takes the
form
\begin{equation}
\Omega^{\mu}_{i}=\Theta_{i}^{\mu}+\frac{3}{2}\,e_{i}^{\mu}\,L_{m}-\frac{1}{2}\,e^{j}_{\alpha}\,\frac{\partial^{2}L_{m}}{\partial
e_{\mu}^{i}\,\partial e_{\alpha}^{j}}
\end{equation}
In this paper, we consider a perfect fluid form for the
stress-energy tensor of the matter as
$\Theta^{i}_{\mu}=(\rho+p)u^{i}\,u_{\mu}-p\,e^{i}_{\mu}$\,, where
$\rho$ is the energy density, $p$ is the pressure and $u_{\mu}$
describes the four-velocity. We also assume that the matter
Lagrangian takes the form $L_{m}=-\rho$ [11] (see [12] for the case
of $L_{m}=-p$). Thus, with these assumptions $\Omega^{\mu}_{i}$ is
rewritten as
\begin{equation}
\Omega^{\mu}_{i}=\Theta_{i}^{\mu}-\frac{3}{2}\,e_{i}^{\mu}\,\rho
\end{equation}
Now in a flat FRW background, $ds^{2}=dt^{2}-a^{2}(t)\,dX^{2}$ with
scale factor $a(t)$, the field equations for $F(T,\Theta)$ gravity
are given by
\begin{equation}
12H^{2}\,F_{T}+F=2\kappa^{2}\,\rho-\rho\,F_{\Theta}
\end{equation}
and
\begin{equation}
48\dot{H}H^{2}F_{TT}-(12H^{2}+4\dot{H})F_{T}-F=2\kappa^{2}p+(5p+6\rho)F_{\Theta}\,.
\end{equation}
where a dot denotes derivative with respect to time. Torsion scalar
as a function of the Hubble parameter $H=\frac{\dot{a}}{a}$ is
expressed by
\begin{equation}
T=-6H^{2}
\end{equation}
Using Eqs. (14) and (15), one can obtain the modified Friedmann
equations as follows
\begin{equation}
3H^{2}=\kappa^{2}(\rho+\rho_{_T}+\rho_{_{(T,\Theta)}})
\end{equation}
and
\begin{equation}
2\dot{H}+3H^{2}=-\kappa^{2}(p+p_{_T}+p_{_{(T,\Theta)}})
\end{equation}
where $\rho_{_T}$ and $p_{_T}$ are energy density and pressure
contribution of torsion scalar, respectively. $\rho_{_{(T,\Theta)}}$
and $p_{_{(T,\Theta)}}$ are the energy density and pressure
contribution of the coupling between torsion and stress-energy
tensor, respectively. These quantities are defined as follows
\begin{equation}
\rho_{_T}=\frac{1}{2F_{T}}(TF_{T}-F)\,,
\end{equation}
\begin{equation}
\rho_{_{(T,\Theta)}}=-\frac{1}{2F_{T}}(\rho\,F_{\Theta})\,,
\end{equation}
\begin{equation}
p_{_T}=\frac{1}{2F_{T}}\Big[F-TF_{T}-48\dot{H}H^{2}F_{TT}\Big]\,,
\end{equation}
and
\begin{equation}
p_{_{(T,\Theta)}}=\frac{1}{2F_{T}}(5p+6\rho)F_{\Theta}\,.
\end{equation}
From Eqs. $(19)-(22)$, we can define gravitationally effective form
of dark energy density $\rho_{DE}=\rho_{_T}+\rho_{_{(T,\Theta)}}$
and pressure $p_{DE}=p_{_T}+p_{_{(T,\Theta)}}$\,, so that equation
of state parameter is defined as
\begin{equation}
\omega_{DE}=\frac{p_{DE}}{\rho_{DE}}=-1+\frac{-48\dot{H}H^{2}F_{TT}+5(p+\rho)F_{\Theta}}{(TF_{T}-F)-\rho
F_{\Theta}}\,.
\end{equation}
\\
In which follows, we want to rewrite the field equations (11) to a
suitable form for our purpose in section 5. To this end, we firstly
multiply $g_{\mu\sigma}e^{i}_{\nu}$ in both sides of (11), so that
the coefficient of the term $F_{T}$ takes the following form
$$e^{-1}e^{i}_{\nu}\partial_{\rho}(ee_{i}^{\alpha}S_{\alpha}^{\,\,\,\mu\rho})-
T^{\lambda}_{\,\,\,\rho\nu}S_{\lambda}^{\,\,\,\mu\rho}$$
$$=\partial_{\rho}S_{\nu}^{\,\,\,\mu\rho}-
\tilde{\Gamma}^{\alpha}_{\,\,\,\nu\rho}S_{\alpha}^{\,\,\,\mu\rho}+\hat{\Gamma}^{\tau}_{\,\,\,\tau\rho}S_{\nu}^{\,\,\,\mu\rho}-
T^{\lambda}_{\,\,\,\rho\nu}S_{\lambda}^{\,\,\,\mu\rho}$$
\begin{equation}
=-\nabla^{\rho}S_{\nu\rho}^{\quad\mu}-K^{\lambda}_{\,\,\,\rho\nu}S_{\,\,\,\lambda}^{\mu\,\,\,\,\rho}
\end{equation}
where we have used the following relation
\begin{equation}
K^{(\mu\nu)\sigma}=T^{\mu(\nu\sigma)}=S^{\mu(\nu\sigma)}=0\,.
\end{equation}
On the other hand, by Eq.(7), the Riemann tensor for the Levi-Civita
connection is written in the following form
\begin{equation}
R^{\rho}_{\,\,\,\mu\lambda\nu}=\partial_{\lambda}\hat{\Gamma}^{\rho}_{\,\,\,\mu\nu}-\partial_{\nu}\hat{\Gamma}^{\rho}_{\,\,\,\mu\lambda}
+\hat{\Gamma}^{\rho}_{\,\,\,\sigma\lambda}\,\hat{\Gamma}^{\sigma}_{\,\,\,\mu\nu}-
\hat{\Gamma}^{\rho}_{\,\,\,\sigma\nu}\,\hat{\Gamma}^{\sigma}_{\,\,\,\mu\lambda}
\end{equation}
then the corresponding Ricci tensor is written as
\begin{equation}
R_{\mu\nu}=\nabla_{\nu}K^{\rho}_{\,\,\,\mu\rho}-\nabla_{\rho}K^{\rho}_{\,\,\,\mu\nu}+K^{\rho}_{\,\,\,\sigma\nu}K^{\sigma}_{\,\,\,\mu\rho}
-K^{\rho}_{\,\,\,\sigma\rho}K^{\sigma}_{\,\,\,\mu\nu}
\end{equation}
By using $K^{\rho}_{\,\,\mu\nu}$ given in Eq. (8) and the relations
(25), and also by considering
$S^{\mu}_{\,\,\rho\mu}=K^{\mu}_{\,\,\rho\mu}=T^{\mu}_{\,\,\mu\rho}$
one obtains [10,13,14]
$$R_{\mu\nu}=-2\nabla^{\rho}S_{\nu\rho\mu}-g_{\mu\nu}\nabla^{\rho}T^{\sigma}_{\,\,\,\rho\sigma}-2K_{\sigma\rho\nu}S^{\rho\sigma}_{\quad\mu}\,,
$$
\begin{equation}
R=-T-2\nabla^{\mu}T^{\nu}_{\,\,\,\mu\nu}\,,
\end{equation}
therefore, one reaches to
\begin{equation}
G_{\mu\nu}-\frac{1}{2}g_{\mu\nu}T=-2\Big(\nabla^{\rho}S_{\nu\rho\mu}+K_{\sigma\rho\nu}S_{\quad\mu}^{\rho\sigma}\Big)
\end{equation}
where $G_{\mu\nu}=R_{\mu\nu}-(1/2)g_{\mu\nu}R$ is the Einstein
tensor. Finally, combining Eqs. (24) and (29), one can rewrite the
field equations for $F(T,\Theta)$ gravity as follows
\begin{equation}
A_{\sigma\nu}F_{T}+B_{\sigma\nu}F_{TT}+\frac{1}{4}g_{\sigma\nu}F(T)-\Omega_{\sigma\nu}F_{\Theta}=\frac{1}{2}\Theta_{\sigma\nu}+\frac{1}{4}\Theta_{\sigma\nu}F_{\Theta}
\end{equation}
where
$$A_{\sigma\nu}=g_{\sigma\mu}e_{\nu}^{i}\Big[e^{-1}\partial_{\rho}(ee^{\alpha}_{i}S_{\alpha}^{\mu\rho})-e^{\lambda}_{i}T^{\alpha}_{\,\,\,\rho\lambda}
S^{\,\,\,\mu\rho}_{\alpha}\Big]$$
$$\quad=-\nabla^{\rho}S_{\nu\rho\sigma}-K_{\lambda\rho\nu}S_{\quad\sigma}^{\rho\lambda}=\frac{1}{2}\Big[G_{\nu\sigma}-(1/2)g_{\nu\sigma}T\Big]
$$
\begin{equation}
B_{\sigma\nu}=S_{\nu\sigma}^{\,\,\,\,\rho}\,\nabla_{\rho}T
\end{equation}
In upcoming sections, we use the trace of Eq. (30) as an independent
relation to simplify the field equation. Since
$A_{\mu}^{\,\,\mu}=-\frac{1}{2}(R+2T)$, the mentioned trace can be
expressed as
\begin{equation}
-\frac{1}{2}(R+2T)F_{T}+BF_{TT}+F(T)+\Omega
F_{\Theta}=\Theta+\frac{1}{4}\Theta F_{\Theta}
\end{equation}
where $B=B_{\mu}^{\,\,\,\mu}$, $\Omega=\Omega_{\mu}^{\,\,\,\mu}$ and
$\Theta=\Theta_{\mu}^{\,\,\,\mu}$ [10].

\section{Perturbations of the flat FRW solutions}

Now we study the homogenous and isotropic perturbations around a
specific cosmological solution for the model described by the action
(1). First, we obtain the perturbed equations for the most general
case. Then as a specific case, the de Sitter solution will be
studied in which follows. Let us assume a general solution in the
FRW cosmological background, which is described by a Hubble
parameter $H=\bar{H}(t)$\,. This solution for a particular
$F(T,\Theta)$ model satisfies equation (14). The matter fluid is
assumed to be in the form of a perfect fluid with a constant
equation of state $p=\omega\,\rho$, in which the matter energy
density $\rho$ satisfies the standard continuity equation
\begin{equation}
\dot{\rho}+3H(1+\omega)\rho=0.
\end{equation}
Then the evolution of the matter energy density is obtained by
solving the continuity equation (33) as follows
\begin{equation}
\hat{\rho}(t)= \rho_{0}\,e^{-3(1+\omega)\int{\bar{H}(t)dt}}
\end{equation}
Where is expressed in term of particular solution $\bar{H}(t)$. In
order to investigate the perturbations around the solutions
$H=\bar{H}(t)$ and energy density (34), small deviations from the
Hubble parameter and the energy density evolution is considered as
[15]
\begin{equation}
H(t)=\bar{H}(t)[1+\delta(t)]\quad ,\quad
\rho(t)=\bar{\rho}(t)[1+\delta_{m}(t)]
\end{equation}
where $\delta(t)$ and $\delta_{m}(t)$ correspond to the isotropic
deviation of the background Hubble parameter and the matter energy
density, respectively. In which follows, to study the behavior of
the perturbations in linear regime, the $F(T,\Theta)$ function is
expanded in the power of $\bar{T}$ and $\bar{\Theta}$ evaluated at
the solution $H=\bar{H}(t)$ as:
\begin{equation}
F(T,\Theta)=\bar{F}+\bar{F}_{T}(T-\bar{T})+\bar{F}_{\Theta}(\Theta-\bar{\Theta})+{\cal{O}}^{2}
\end{equation}
where a bar indicates the value of $F(T,\Theta)$ function and its
derivatives evaluated at $T=\bar{T}$ and $\Theta=\bar{\Theta}$\,. By
substituting Eqs. (35) and (36) into the Friedmann equation (14),
one can obtain an expression for the perturbations $\delta(t)$ in
linear approximation as follows
\begin{equation}
12\bar{H}^{2}\bar{F}_{T}\,\delta(t)+\Big[\bar{F}+(\Theta-\bar{\Theta})\bar{F}_{\Theta}-(2\kappa^{2}-\bar{F}_{\Theta})\bar{\rho}_{m}\Big]=
(2\kappa^{2}-\bar{F}_{\Theta})\,\bar{\rho}_{m}\,\delta_{m}(t)
\end{equation}
It seems that Eq. (37) is an algebraic equation for $\delta(t)$, but
since $\Theta$ as trace of the stress-energy tensor itself is
expressed in terms of $H$ and $\dot{H}$, one should find a
differential equation for $\delta(t)$\,. For this purpose, we
firstly contract the field equations (11) by $e_{\mu}^{i}$ to find
\begin{equation}
\Theta=\frac{4}{2+F_{\Theta}}\Bigg[e^{-1}F_{T}\Big[\partial_{\rho}(eS_{\mu}^{\,\,\mu\rho})-e\,(\partial_{\rho}e^{i}_{\mu})\,S_{i}^{\,\,\mu\rho}\Big]+TF_{T}
+S_{\mu}^{\,\,\mu\rho}\partial_{\rho}T\,F_{TT}+F-\Omega_{\mu}^{\mu}\,F_{\Theta}\Bigg]
\end{equation}
It is easy to show that
\begin{equation}
\partial_{\rho}(eS_{\mu}^{\,\,\mu\rho})=3e\,(\dot{H}+H^{2})\,,\quad
(\partial_{\rho}e^{i}_{\mu})\,S_{i}^{\,\,\mu\rho}=3H^{2}\,,\quad
S_{\mu}^{\,\,\mu\rho}\partial_{\rho}T=3H\dot{H}\,.
\end{equation}
Also by using expression (13) and
$\Omega_{\mu}^{\mu}=e_{\mu}^{i}\,\Omega_{i}^{\mu}$\,, one obtains
\begin{equation}
\Omega_{\mu}^{\mu}=\Theta-6\rho_{m}\,.
\end{equation}
So, the expression for $\Theta$ can be deduced as follows
\begin{equation}
\Theta=\frac{12}{2+5F_{\Theta}}\Big[(\dot{H}-2H^{2})F_{T}+2\rho_{m}\,F_{\Theta}
+H\dot{H}\,F_{TT}+\frac{1}{3}F\Big]
\end{equation}
Now one can substitute Eqs. (35) and (36) into the expression for
$\Theta$ and then $\Theta$ in Eq. (37) in order to get the
corresponding differential equation for $\delta(t)$ as follows
\begin{equation}
C_{2}\,\dot{\delta}(t)+C_{1}\,\delta(t)+C_{0}=C_{m}\,\delta_{m}(t)\,,
\end{equation}
where $C_{0\,,1\,,2}$ and $C_{m}$ depend on the $F(T,\Theta)$ and
its derivatives which are explicitly written in the following
\begin{equation}
C_{2}=\frac{12\bar{H}}{2+\bar{F}_{\Theta}}\,\bar{F}_{\Theta}\,\bar{F}_{T}+\bar{H}^{2}\,\bar{F}_{TT}\,,
\end{equation}
\begin{equation}
C_{1}=12\,\bar{H}^{2}\bar{F}_{T}+\frac{12\bar{F}_{\Theta}}{2+\bar{F}_{\Theta}}\,\Big(\dot{\bar{H}}-8\bar{H}^{2}\Big)\bar{F}_{T}
+2\bar{H}\dot{\bar{H}}\,\bar{F}_{TT}\,,
\end{equation}
\begin{equation}
C_{0}=\bar{F}-(2\kappa^{2}-\bar{F}_{\Theta})\bar{\rho}_{m}-\bar{\Theta}\bar{F}_{\Theta}
+\frac{12\bar{F}_{\Theta}}{2+\bar{F}_{\Theta}}\,\Bigg[\bar{H}\dot{\bar{H}}\,\bar{F}_{TT}+
(\dot{\bar{H}}-2\bar{H}^{2})\bar{F}_{T}+2\bar{\rho}_{m}\,\bar{F}_{\Theta}-\bar{\Theta}\bar{F}_{\Theta}+\frac{1}{3}\bar{F}\Bigg]
\,,
\end{equation}
and
\begin{equation}
C_{m}=\bar{\rho}_{m}\Big(2\kappa^{2}-\bar{F}_{\Theta}-\frac{24(\bar{F}_{\Theta})^{2}}{2+\bar{F}_{\Theta}}\Big)\,,
\end{equation}
Also, there is another perturbed equation which is obtained from the
matter continuity equation (33) and perturbed expressions (35), as
follows
\begin{equation}
\dot{\delta}_{m}(t)+3\bar{H}(t)\,\delta(t)=0
\end{equation}
In general relativity, the stability equation (42) is reduced to an
algebraic relation between geometrical and the matter perturbations.
For higher order theories of gravity, the evolution of the
perturbations is in general determined by a system composed of
ordinary differential equations (42) and (47). Equation (42) is a
non-homogeneous and linear first order differential equation. To
solve this differential equation, one firstly rewrites it in the
standard form and then finds an integrating factor. Multiplying the
standard equation by integrating factor, $\delta(t)$ can be obtained
as
\begin{equation}
\delta(t)=\Bigg[\int{\frac{e^{\frac{C_{1}}{C_{2}}t}[C_{m}\delta_{m}(t)-C_{0}]}{C_{2}}dt}\Bigg]e^{-\frac{C_{1}}{C_{2}}t}
\end{equation}
Hence, for a FRW cosmological solution, the stability of the
solution can be investigated in the framework of $F(T,\Theta)$
gravity by solving the equations (42) and (47). In the next section,
we will illustrate the previous discussions by considering theories
which include the de Sitter solution as the simplest cosmological
solution.

\section{The stability of the de Sitter solution}

The de Sitter solution is one of the simplest cosmological solutions
which can realize the late-time accelerated phase of the universe
expansion as well as the inflationary epoch. On the other hand, the
existence of a \emph{stable de Sitter solution} helps the theory to
be cosmologically viable. Therefore, we study the stability of the
de Sitter solution,
\begin{equation}
H(t)=H_{(0)}\quad , \quad a(t)=a_{0}e^{H_{(0)}t}
\end{equation}
where $H_{(0)}$ is a constant. Since the de Sitter solution is a
vacuum solution, the perturbations is depend on only the underlying
gravitational theory. According to the differential equation for the
perturbations, Eq. (42), now we have
\begin{equation}
C_{2}^{(0)}\,\dot{\delta}(t)+C_{1}^{(0)}\,\delta(t)+C_{0}^{(0)}=0
\end{equation}
with
\begin{equation}
C_{2}^{(0)}=F^{(0)}_{\Theta}\Big(12F^{(0)}_{T}+F^{(0)}_{TT}\Big)+2F^{(0)}_{TT}\,,\,\,\,
C_{1}^{(0)}=12F^{(0)}_{T}\Big(2-7F^{(0)}_{\Theta}\Big)\,,\,\,\,
C_{0}^{(0)}=\frac{F^{(0)}}{H_{(0)}^{2}}(6+F^{(0)}_{\Theta})-24F^{(0)}F^{(0)}_{\Theta}
\end{equation}
where the notation $(0)$ indicates the value of each function
evaluated in the de Sitter phase $T=T_{(0)}$ and
$\Theta=\Theta_{(0)}$. The general solution of equation (50)
demonstrates the dynamical behavior of the gravitational
perturbations as follows
\begin{equation}
\delta(t)=-\frac{C_{0}^{(0)}}{{C}_{1}^{(0)}}+\alpha
e^{-\frac{C_{1}^{(0)}}{{C}_{2}^{(0)}}\,t}
\end{equation}
where $\alpha$ is an arbitrary integration constant. As we see, the
growth of the gravitational perturbations tends to the constant
value $-\frac{C_{0}^{(0)}}{{C}_{1}^{(0)}}$ with the stability
condition:
\begin{equation}
\frac{C_{1}^{(0)}}{{C}_{2}^{(0)}}=\frac{12F^{(0)}_{T}(2-7F^{(0)}_{\Theta})}{12F^{(0)}_{\Theta}F^{(0)}_{T}+2F^{(0)}_{TT}+F^{(0)}_{\Theta}F^{(0)}_{TT}}>0
\end{equation}
As we see the stability of the de Sitter solution depends explicitly
on the values of the function $F(T,\Theta)$ and its derivatives at
$T_{(0)}$ and $\Theta_{(0)}$.

In order to display the previous calculations, we consider the
$F(T,\Theta)$ function as follows (see for instance [15])
\begin{equation}
F(T,\Theta)=k_{1}\,T+k_{2}\,T^{m}\Theta^{n}
\end{equation}
where $k_{1}$ and $k_{2}$ are positive or negative coupling
constants. Now one can easily solve Eq. (14) to find
\begin{equation}
\Theta_{(0)}^{n}=\frac{k_{1}}{(1-2m)k_{2}}T_{(0)}^{^{1-m}}
\end{equation}
Then by using Eqs. (16) and (41) (in the linear regime), we obtain
\begin{equation}
\Theta_{(0)}=\frac{k_{1}}{2(1-2m)}T_{(0)}\Big[12(1-m)-5n\Big]
\end{equation}
Now, the combination of the last two equations gives the following
de Sitter solution
\begin{equation}
H_{(0)}=\Bigg[\Big(-6\Big)^{m+n-1}\Big(\frac{k_{1}}{1-2m}\Big)^{n-1}\Big(6[1-m]-\frac{5}{2}n\Big)^{n}\,k_{2}\Bigg]^{\frac{1}{2(1-m-n)}}
\end{equation}
As a specific example, we consider the case with $m=\frac{2}{3}$ and
$n=1$\,. Thus the de Sitter solution takes the following form
\begin{equation}
H_{(0)}=\Big(\,\frac{-2}{9\,k_{2}^{3}}\,\Big)^{\frac{1}{4}}
\end{equation}
As we see, in this model there is a determinate district for $k_{2}$
and it is that for a universe with a well-defined de Sitter
expansion, $k_{2}$ must be negative. Now, the stability condition
(53) by using (56) gives
\begin{equation}
\frac{36\,(1+k_{2}T_{0}^{\frac{2}{3}})(2-7k_{2}T_{0}^{\frac{2}{3}})}
{36k_{2}T_{0}^{\frac{2}{3}}\,(1+k_{2}T_{0}^{\frac{2}{3}})-2k_{2}T_{0}^{-\frac{1}{3}}-k_{2}^{2}T_{0}^{\frac{1}{3}}}>0
\end{equation}
Substituting (16) and (58) into (59), the stability condition is
rewritten as
$$\frac{-576}{144+\sqrt{2}\,(-k_{2})^{\frac{7}{6}}+\sqrt{2}(-k_{2})^{\frac{5}{6}}}>0\,.$$
It is clear that this inequality is incorrect. In other words, in
the model with $m=\frac{3}{2}$ and $n=1$ the perturbation grows
exponentially and the de Sitter solution becomes unstable. So, it is
not cosmologically a viable model. The other example is a model with
$m=\frac{6}{7}$ and $n=1$ (we note that these choices are restricted
from the energy conditions viewpoint as we see later). In this case,
the de Sitter solution takes the form
$H_{(0)}=(-32\,k_{2}^{7}\,)^{-\frac{1}{12}}$\,. Again for a universe
with a well-defined de Sitter expansion, $k_{2}$ must be negative.
Here the stability condition (53) by using (56) and (58) takes the
following form
\begin{equation}
\frac{210}{0.63\,(-k_{2}^{7})^{\frac{1}{6}}-25.6}>0
\end{equation}
Thus the stability condition reduces to
\begin{equation}
k_{2}< -0.76
\end{equation}
Note that for the models with $n=1$, the stability condition has no
dependence to the parameter $k_{1}$\,. Generally, to recover the
teleparallel equivalent of the General Relativity, $k_{1}$ should be
positive.\\
In the upcoming section, we discuss the energy conditions in the
general $F(T,\Theta)$ theories, then we specify a kind of the
$F(T,\Theta)$ function in the spirit of (54) and finally we
investigate whether the energy conditions can be satisfied in the
context of the constraint (61) or not.

\section{Energy conditions}
The Raychaudhuri equation is origin of the strong and null energy
conditions together with the requirement that gravity is attractive
for the space time manifold that is endowed by a metric
$g_{\mu\nu}$. For a congruence of timelike geodesics with tangent
vector field $u^{\mu}$, Raychaudhuri equation as the temporal
variation of expansion $\theta$ [16] is defined as follows
\begin{equation}
\frac{d\theta}{d\tau}=-\frac{1}{3}\theta^{2}-\sigma_{\mu\nu}\sigma^{\mu\nu}+
\omega_{\mu\nu}\omega^{\mu\nu}-R_{\mu\nu}u^{\mu}u^{\nu}
\end{equation}
where $\theta$ is expansion parameter, $\sigma^{\mu\nu}$ and
$\omega_{\mu\nu}$ are, respectively, shear and rotation associated
with the congruence defined by the vector field $u^{\mu}$\,. In the
case of null vector field $n^{\mu}$ the temporal version of the
expansion is given by
\begin{equation}
\frac{d\theta}{d\tau}=-\frac{1}{2}\theta^{2}-\sigma_{\mu\nu}\sigma^{\mu\nu}+
\omega_{\mu\nu}\omega^{\mu\nu}-R_{\mu\nu}n^{\mu}n^{\nu}
\end{equation}
Note that the Raychaudhuri equation is a purely geometric equation
and hence, it is not restricted to a specific theory of gravitation.
Since the shear tensor is purely spatial
$\sigma_{\mu\nu}\sigma^{\mu\nu}>0$, thus, for any hypersurface of
orthogonal congruence ($\omega_{\mu\nu}=0$), the conditions for
gravity to be attractive, become
\begin{equation}
\textbf{SEC}:\quad\quad\quad\quad  R_{\mu\nu}u^{\mu}u^{\nu}\geq 0
\end{equation}
\begin{equation}
\textbf{NEC}:\quad\quad\quad\quad  R_{\mu\nu}n^{\mu}n^{\nu}\geq 0
\end{equation}
By using the field equations of any gravitational theory, one can
relate the Ricci tensor to the energy-momentum tensor
$\Theta_{\mu\nu}$. Thus, the combination of the field equations and
Raychaudhuri equation sets a series of physical conditions for the
energy-momentum tensor. By employing (64) and (65) in the general
relativity framework, one can restrict the energy-momentum tensor as
follows
\begin{equation}
R_{\mu\nu}u^{\mu}u^{\nu}=\Big(\Theta_{\mu\nu}-\frac{1}{2}g_{\mu\nu}\Theta\Big)u^{\mu}u^{\nu}\geq0
\end{equation}
and
\begin{equation}
R_{\mu\nu}n^{\mu}n^{\nu}=\Theta_{\mu\nu}n^{\mu}n^{\nu}\geq0
\end{equation}
where for a perfect fluid with density $\rho$ and pressure $p$\,,
this expression reduces to the well-known forms of the SEC and NEC
in general relativity:
\begin{equation}
\rho+3p\geq0 \quad\quad and \quad\quad\rho+p\geq0
\end{equation}
Note that in the inequalities (66) and (67) we have set
$\kappa^{2}=1$. In which follows we continue to use this convention.

\subsection{The energy condition in $F(T,\Theta)$ gravity}

The Raychaudhuri equation together with attractor character of the
gravitational interaction have led us to Eqs. (66) and (67). These
relations hold for any theory of gravity. In which follows, we apply
this approach to drive the strong energy condition (SEC) and null
energy condition (NEC) in $F(T,\Theta)$ gravity. First we rewrite
the field equations (30) as follows
\begin{equation}
G_{\mu\nu}=\frac{1}{F_{T}}\Big(\Theta_{\mu\nu}-2B_{\mu\nu}F_{TT}+\frac{1}{2}[TF_{T}-F]g_{\mu\nu}+
2F_{\Theta}[\Omega_{\mu\nu}+\frac{1}{4}\Theta_{\mu\nu}]\Big)
\end{equation}
From this equation and the trace of the field equations, Eq. (32),
we have
\begin{equation}
R_{\mu\nu}={\cal T}_{\mu\nu}-\frac{1}{2}{\cal T}g_{\mu\nu}
\end{equation}
where
\begin{equation}
{\cal
T}_{\mu\nu}=\frac{1}{F_{T}}\Big[\Theta_{\mu\nu}-2B_{\mu\nu}F_{TT}+2F_{\Theta}(\Omega_{\mu\nu}+\frac{1}{4}\Theta_{\mu\nu})\Big]
\end{equation}
\begin{equation}
{\cal
T}=\frac{1}{F_{T}}\Big[\Theta+TF_{T}-F-2BF_{TT}-2F_{\Theta}(\Omega-\frac{1}{4}\Theta)\Big]
\end{equation}
Now in a FRW background, from Eqs. (6) and (9) along with Eq. (31),
we obtain
\begin{equation}
\quad\,\,A_{00}=3H^{2}\,,\quad\quad\quad
A_{ij}=-a^{2}(3H^{2}+\dot{H})\delta_{ij}
\end{equation}
\begin{equation}
B_{ij}=12a^{2}H^{2}\dot{H}\delta_{ij}\,,\quad\quad\quad
B=-36H^{2}\dot{H}
\end{equation}
For a perfect fluid of density $\rho$ and pressure $p$,
$\Theta_{\mu\nu}=e^{\alpha}_{i}\,g_{\alpha\nu}\,\Theta^{i}_{\mu}$\,,
and by taking $u_{\mu}=(1,0,0,0)$ and $n_{\mu}=(1,a,0,0)$, we obtain
the ${\cal T}_{\mu\nu}$ and its trace ${\cal T}$ as follows
\begin{equation}
{\cal
T}_{00}=\frac{1}{F_{T}}\Big(1-\frac{F_{\Theta}}{4}\Big)\,\rho\,,\quad\quad
{\cal
T}_{ij}=\frac{a^{2}}{F_{T}}\Big(p-24H^{2}\dot{H}F_{TT}+\frac{1}{2}(5p+6\rho)F_{\Theta}\Big)\delta_{ij}
\end{equation}
and
\begin{equation}
{\cal
T}=\frac{1}{F_{T}}\Big[\rho-3p+TF_{T}-F+72H^{2}\dot{H}F_{TT}+\frac{1}{2}F_{\Theta}(21\rho+9p)\Big]
\end{equation}
Here we can use equations (64) and (65) together with Eqs. (75) and
(76), for a general $F(T,\Theta)$ gravity, to achieve the strong and
null energy conditions respectively as
\begin{equation}
\textbf{SEC}:\quad\quad
\frac{1}{2F_{T}}\Big(\rho+3p+F-TF_{T}-72H^{2}\dot{H}F_{TT}-\frac{1}{2}(23\rho+9p)F_{\Theta}\Big)\geq
0
\end{equation}
\begin{equation}
\textbf{NEC}:\quad\quad
\frac{1}{F_{T}}\Big(\rho+p-24H^{2}\dot{H}F_{TT}+\frac{3}{2}(\rho+p)F_{\Theta}\Big)\geq
0\quad\quad\quad\quad\quad\quad\quad\quad
\end{equation}
As one may expect, the energy conditions in the general relativity
framework \emph{i.e.} Eq. (68), can be recovered as a particular
case of SEC and NEC in the context of $F(T,\Theta)$ gravity if we
set $F(T,\Theta)=T$\,.\\ By defining an effective energy-momentum
tensor in the context of $F(T,\Theta)$ gravity, SEC and NEC can also
be recasted in the form of that of GR ( $\rho^{eff}+3p^{eff}\geq0$
and $\rho^{eff}+p^{eff}\geq0$\,, respectively). In this respect we
can drive weak energy condition (WEC) and dominant energy condition
(DEC). The effective energy-momentum tensor in the framework of
$F(T,\Theta)$ gravity is defined as follows (similar to $F(T)$
gravity in Ref. [14])
\begin{equation}
\Theta^{eff}_{\mu\nu}=\frac{1}{F_{T}}\Big[\Theta_{\mu\nu}-2B_{\mu\nu}F_{TT}+\frac{1}{2}(TF_{T}-F)g_{\mu\nu}+
2F_{\Theta}(\Omega_{\mu\nu}+\frac{1}{4}\Theta_{\mu\nu})\Big]
\end{equation}
$\rho^{eff}$ and $p^{eff}$ can be derived via the effective
energy-momentum tensor by the following definitions
\begin{equation}
\rho^{eff}=g^{00}\Theta^{eff}_{00}\,, \quad\quad
p^{eff}=-\frac{1}{3}g^{ij}\Theta^{eff}_{ij}
\end{equation}
Thus, using the effective energy-momentum tensor approach, the weak
energy condition (WEC) in $F(T,\Theta)$ gravity ($\rho^{eff}\geq0$)
is written as
\begin{equation}
\textbf{WEC} : \quad\quad
\frac{1}{F_{T}}\Big[\rho+\frac{1}{2}(TF_{T}-F)-\frac{1}{2}\rho
F_{\Theta}\Big]\geq0
\end{equation}
Finally one can write the dominant energy condition
($\rho^{eff}>|p^{eff}|\geq0$ ) as follows
\begin{equation}
\textbf{DEC} :\quad\quad
\frac{1}{F_{T}}\Big[\rho-p+(TF_{T}-F)+24H^{2}\dot{H}F_{TT}-\frac{1}{2}(7\rho+5p)
F_{\Theta}\Big]\geq0
\end{equation}
In the next section, we test one of the $F(T,\Theta)$ models in the
context of the energy conditions we derived. In this way we obtain a
constraint on the parametric space of the model.

\subsection{Constraining $F(T,\Theta)$ models from energy conditions}

We firstly list the energy conditions in terms of the
phenomenological parameter of deceleration
$q=-\frac{\ddot{a}}{a}H^{-2}=-(1+\frac{\dot{H}}{H^{2}})$. The
positivity of the Newtonian gravitational constant requires also the
constraint $F_{T}>0$. With these notifications, the energy
conditions are rewritten as follows
\begin{equation}
\textbf{WEC}:\quad\quad
2\rho_{0}+T_{0}F_{T_{0}}-F_{0}-\rho_{0}F_{\Theta_{0}}\geq0\quad\quad\quad\quad\quad\quad\quad\quad\quad\quad\quad\quad\quad\quad\quad\quad\quad\,\,
\end{equation}
\begin{equation}
\textbf{NEC}:\quad\quad
\rho_{0}+p_{0}+24H_{0}^{4}(1+q_{0})F_{T_{0}T_{0}}+\frac{3}{2}(\rho_{0}+p_{0})F_{\Theta_{0}}\geq
0\quad\quad\quad\quad\quad\quad\quad\quad\quad
\end{equation}
\begin{equation}
\textbf{SEC}:\quad\quad
\rho_{0}+3p_{0}+F_{0}-T_{0}F_{T_{0}}+72H_{0}^{4}(1+q_{0})F_{T_{0}T_{0}}-\frac{1}{2}(23\rho_{0}+9p_{0})F_{\Theta_{0}}\geq0\quad
\end{equation}
\begin{equation}
\textbf{DEC}:\quad\quad
\rho_{0}-p_{0}+T_{0}F_{T_{0}}-F_{0}-24H_{0}^{4}(1+q_{0})F_{T_{0}T_{0}}-\frac{1}{2}(7\rho_{0}+5p_{0})
F_{\Theta_{0}}\geq0\quad\quad
\end{equation}
We note that all the above conditions depend on the present value of
pressure $p_{0}$, so for simplicity we assume $p_{0}=0$\,.\\
Then we should adopt a specific function for $F(T,\Theta)$ to obtain
the constraints on the parametric space of the considered model from
the point of view of the energy conditions. On the other hand, we
know that in order for a theoretical model to be cosmologically
viable, it should satisfy at least the weak energy condition. This
leads us to the mentioned constraints on parametric space of the
model. Here we again consider
$F(T,\Theta)=k_{1}T+k_{2}T^{m}\Theta^{n}$ as our background
gravitational model. The weak energy condition together with Eqs.
(55) and (56) is satisfied by
\begin{equation}
\frac{m-1}{1-2m}\,k_{1}\leq
\,\Omega_{m_{0}}\,\Big(1+\frac{n}{12m+5n-12}\Big)
\end{equation}
By restricting the parameter $m$ values, one can constrain the
parameter $k_{1}$. Also to recover the teleparallel equivalent of
the General Relativity, $k_{1}$ should be positive. Now one can
obtain three ranges for $m$ as $\frac{1}{2}< m < 1$\,, $m>1$\, and
$m<\frac{1}{2}$ in which the
constraint (87) is rewritten as follows\\

\textbf{1. The case with $m<\frac{1}{2}$ and $1<m$\,:}
\begin{equation}
k_{1}\geq \frac{1-2m}{m-1}
\,\Omega_{m_{0}}\,\Big(1+\frac{n}{12m+5n-12}\Big)
\end{equation}
Here by considering the condition for recovery of General
Relativity, that is, $k_{1}>0$, we are led to the other constraint
on the parameter $n$ and $m$ as
$$5\leq\frac{5n}{2(1-m)}<6\,.$$

\textbf{2. The case with $\frac{1}{2}< m < 1$\,:}
\begin{equation}
k_{1}\leq \frac{1-2m}{m-1}
\,\Omega_{m_{0}}\,\Big(1+\frac{n}{12m+5n-12}\Big)
\end{equation}
Now the mentioned condition, $k_{1}>0$, leads us to the following
constraint:
$$n<2(1-m) \,\,,\quad\quad  n>2.4(1-m)$$ So, in the model
with $m=\frac{2}{3}$ and $n=1$ which is considered in section 4, the
weak energy condition can be realized with condition $k_{1}\leq
2\Omega_{m_{0}}$, but the de Sitter solution is unstable in this
case. Nevertheless, we could find the suitable values for $m$ and
$n$ for which the weak energy condition can be realized as well as
the stable de Sitter solution. This can be done if we set
$m=\frac{6}{7}$ and $n=1$ for instance. In this case the weak energy
condition holds if $k_{1}\leq \frac{150}{23}\Omega_{m_{0}}$ and the
de Sitter
phase is stable.\\
Also one can investigate consistency of the null energy condition in
the de Sitter phase. Note that one should set the value of
$q_{0}=-1$ for the de Sitter phase, so that the coefficient of the
term $F_{T_{0}T_{0}}$ in Eq. (85) vanishes. The NEC in the de Sitter
phase imposes a constraint on the parameters $m$ and $n$ as follows
\begin{equation}
\frac{n}{12(m-1)+5n}\leq \frac{1}{3}
\end{equation}
On the other hand, the constraints of WEC on the $m$ and $n$ (along
with the positivity of $k_{1}$) which have already been mentioned,
can be used to obtain a more restricted ranges of the parameters $m$
and $n$\,. For example, in the second case in which $\frac{1}{2}< m
< 1$, the NEC in the de Sitter solution imposes the following
constraints
\begin{equation}
n<2(1-m)\,, \quad\quad\quad n>6(1-m)
\end{equation}
So, in the model with $m=\frac{2}{3}$ and $n=1$, in spite of the
realization of the WEC with condition $k_{1}\leq 2\Omega_{m_{0}}$,
the null energy condition can not be satisfied. While in the model
with $m=\frac{6}{7}$ and $n=1$ in the de Sitter phase, both of the
WEC (with $k_{1}\leq \frac{150}{23}\Omega_{m_{0}}$\,) and NEC are
realized as well as the stable de Sitter solution. Thus, the later
model is cosmologically viable.

\section{Conclusion}

In this work we discussed the cosmological viability of an
alternative gravitational theory, namely, the modified $F(T,\Theta)$
gravity, where $T$ is the Torsion scalar and $\Theta$ is the trace
of the energy-momentum tensor. The viability of the model is based
on the existence of a stable de Sitter solution and the realization
of all the energy conditions or at least some of them. In a
perturbational approach, we have obtained a differential equation
for $\delta(t)$\,. As a special case, we analyzed the differential
equation for the de Sitter solution and we obtained a condition for
the stability of this solution. Then we focused on the case where
the algebraic function $F(T,\Theta)$ is cast into
$F(T,\Theta)=k_{1}T+k_{2}T^{m}\,\Theta^{n}$, where $k_{1}$, $k_{2}$,
$m$ and $n$ are input parameters. We firstly adopted the case with
$m=\frac{2}{3}$ and $n=1$ and we have shown that the perturbations
in the model grow with time exponentially. Then we considered the
other case with $m=\frac{6}{7}$ and $n=1$. This model realizes a
stable de Sitter phase with the condition $k_{2}<-0.76$. Note that
for simplicity we have adopted the value $n=1$, because by this
choice one gets rid of the dependence of the stability condition to
the parameter $k_{1}$\,. Finally we investigated the energy
conditions in the $F(T,\Theta)$ models. We focused on the fact that
WEC is the main condition for the cosmological viability of the
theory to obtain a constraint on the parameters $m$\,, $n$ and
$k_{1}$. Then by assuming that the parameter $k_{1}$ should be
positive to recover the teleparallel equivalent of the General
Relativity, we achieved the more restricted parametric space for $m$
and $n$\,. In the next step, the adopted values for $m$ and $n$ (in
the stability discussion) are applied. We have shown that the case
with $m=\frac{2}{3}$\,,  $n=1$ and the other case with
$m=\frac{6}{7}$ and $n=1$ can realize the WEC along with
$k_{1}<2\Omega_{m_{0}}$ and $k_{1}\leq \frac{150}{23}
\Omega_{m_{0}}$\,, respectively. In the last step we considered the
cosmological viability of the model from the point of view of the
NEC\,. Since the purpose of our study was the comparison of the
energy conditions with the stability of the de Sitter phase, we
considered NEC at $q=-1$ (in the de Sitter solution). Here we
obtained the more complete constraint on the $m$ and $n$\,, so that
it entails both WEC and NEC\,. As we saw, the case with
$m=\frac{6}{7}$ and $n=1$ realizes NEC too and is cosmologically a
viable gravitational theory.

\end{document}